**Article type: Research news**

**Residual strain and stress in biocrystals**

*Eva Seknazi, and Boaz Pokroy\**

E. Seknazi, Prof. B. Pokroy
Department of Materials Science and Engineering and the Russell Berrie Nanotechnology Institute, Technion – Israel Institute of Technology, 32000 Haifa, Israel
E-mail: bpokroy@tx.technion.ac.il

Keywords: biominerals, residual strains, residual stresses, toughening strategies

The development of residual strains within a material is a valuable engineering technique for increasing the material's strength and toughness. Residual strains occur naturally in some biominerals and are an important feature that has recently been highlighted in biomineral studies. Here, we review manifestations of internal residual strains detected in biominerals. We describe the mechanisms by which they develop as well as their impact on the biominerals' mechanical properties. The question as to whether they can be utilized in multi-scale strengthening and toughening strategies for biominerals is discussed.

**1. Introduction**

Biominerals consist mainly of brittle ceramic materials, and small amounts of soft organic matter.[1, 2] To compensate for their small variety of available constituents they use elaborate structural strategies, enabling them to achieve mechanical properties far superior to those of their non-biological counterparts. These strategies rely among other features on the combined use of organic matter and minerals in multi-scale hierarchically organized architectures.[1-6] Often the outcome is the deflection and arrest of crack propagation, resulting in a toughened material.[6] This review focuses on another structural strategy, which occurs at the smallest scale of the biomineral—the exploitation of residual strain and stresses within the crystals.

Internal residual stresses are the stresses experienced by the material without any external stress application. The existence of internal residual stresses can result in strengthening of the material's resistance to externally applied stress in the opposite direction, which would have to overcome the internal stress of the material. Thus, compressive residual stresses will result in increased tensile strength, and tensile residual stress will result in increased compressive strength. In addition, compressive or tensile local stress fields in a crystal can interact with cracks and deflect them, resulting in a toughened material. Similarly, local stress fields in a crystal can interact with dislocation or twin stress fields and impede their motion, resulting in a material with increased hardness. Furthermore, since cracks open as a result of tensile stresses, the existence of internal compressive stresses will delay the opening of cracks until additional tensile stress is applied to overcome the internal compressive stress, resulting in toughening of the material. The intentional integration of residual compressive stresses within a brittle material is a widespread engineering procedure used to enhance the material's mechanical properties. This procedure is used, for example, in the toughened glass that we use for products such as smartphones, or in prestressed concrete for building. Likewise, their presence in various biominerals may serve to strengthen and toughen the material. Therefore, the understanding of how such stresses are implanted in the mineral are of interest for the design of new engineering strengthening and toughening strategies.

In this paper, we review key occurrences of both tensile and compressive residual internal stresses and strains reported in biominerals. We differentiate between the cases of residual stress caused by intracrystalline organic and inorganic inclusions, and those caused by organic or inorganic interactions at intercrystalline interfaces. The circumstances of their occurrence and their effect on the biominerals' mechanical properties are described.

## 2. Residual stresses and strains caused by intracrystalline inclusions

### 2.1. Residual stresses and strains caused by organic inclusions

One of the most characteristic features of biogenic crystals is the presence of organic molecules that interact with the crystals. The major role of intercrystalline organic macromolecules in biogenic crystals is well studied and accepted. Perhaps the best-known example of this is the brick and mortar structure of nacre, in which mineral tiles are piled up on each other and separated by organic layers that serve as barriers to crack propagation and allow the tiles to slide on each other.[1] Organic molecules are also present within the crystals, as intracrystalline inclusions.[7-10] The existence of organic intracrystalline inclusions has proved to be a common phenomenon in biogenic calcium carbonate crystals and to be responsible for lattice distortions not occurring in their geologic counterparts. Lattice distortions have frequently been reported in aragonite[11-19] and calcite[20-23] mollusk shells. Such strains were indeed detected when we carried out precise measurements on these lattice parameters by means of synchrotron X-ray diffraction on powdered bleached crystals. We found that upon heating the lattice strains relaxed, returning to the established parameters of the geological crystals. The relaxation occurred at temperatures that coincided with mass loss and decomposition of organic matter, as signaled by thermal gravimetric analysis and mass spectrometric measurements. Moreover, relaxation of the lattice parameters was accompanied by a broadening of the diffraction peaks, corresponding to the reduction in size of the coherently scattered crystallites. Taken together, these observations indicated that occluded organic molecules are present in the host lattice and lead to lattice strains. Their decomposition relaxes the lattice and dismantles the room temperature well-arranged organization of the organic/inorganic biocomposite. The measured lattice distortions caused by organic molecules were anisotropic. In the case of the orthorhombic aragonite, intracrystalline organic inclusions caused an increase in the *a*- and *c*-lattice parameters while the *b*-lattice parameter decreased. In the case of rhombohedral calcite, both the *a*- and *c*-lattice parameters increased. In both biogenic calcite and aragonite, the highest distortion was obtained for the *c*-lattice parameter and reached values up to 0.2%. These lattice strains corresponded to stress components reaching 182 MPa in the orthorhombic structure of aragonite and 220 MPa for the rhombohedral structure of calcite.[8]

Moreover, synthetic calcite grown in the presence of extracted macro molecules,[24-26] single amino acid,[27, 28] copolymer micelles[29] or even 2D graphene oxide sheets,[30] have been synthesized and were found to display lattice distortions that were sometimes comparable to those detected in biogenic crystals. Taken together, these studies corroborate the ability of calcite to host organic molecules and sustain high lattice distortions.

With regard to the mechanical properties, tensile stresses present in the crystals increase the compressive strength of the minerals, making it harder for predators to crush them.[9] Moreover, as mentioned earlier, the stress fields around the molecules can interact with the stress fields of cracks or dislocations and inhibit their motion, resulting in enhanced toughness and hardness. In the case of synthetic calcite hosting organic molecules, the increase in hardness was observed by means of nano-indentation measurements.[27, 29] However, the hardening mechanism was interpreted as being attributable to the direct blocking of dislocations by the molecules themselves rather than to the strain fields they cause.[27]

The origin of these tensile distortions is believed to be related to the crystallization pathways of the biogenic crystals. Biogenic calcite and aragonite indeed often originate from the crystallization of amorphous calcium carbonate,[31-33] which is associated with water loss and shrinkage. Organic molecules present in the amorphous phase can participate in stabilizing this phase and dictating the crystal selection and shape.[33] Thus, the presence of organic molecules within the amorphous phase would exert resistance to the shrinkage during crystallization, resulting in the distortions observed in the crystalline phase. It therefore remains an open question whether the tensile strains on the biogenic crystals are part of the organism's strategies, or are merely a side effect of the utilization of amorphous calcium carbonate as a precursor

**2.2. Residual stresses and strains caused by inorganic inclusions**

Calcitic marine biominerals usually contain a certain amount of magnesium. In such cases the magnesium atoms can substitute for calcium atoms in the calcite. The amount of magnesium varies between biominerals,[34] from less than 1 mol% to 45 mol%; the latter amount is found in sea urchin teeth, the richest magnesium biogenic calcite found to date.[35, 36] Since the Ca ion is almost twice the size of the Mg ion, Mg substitution causes shortening of interatomic distances, high internal strain, and lattice distortions.[37] Experimental[35, 38] and computational studies[39, 40] have shown that the homogeneous incorporation of magnesium into calcite enhances the host crystal hardness and stiffness. However, Mg substitution in biogenic calcite is not entirely homogeneous, and biogenic Mg-calcite can display extensive Mg distribution that most probably plays a role in the superior mechanical properties of biominerals.[36, 41, 42]

A recent study by our group of the microscopic optic lenses of the brittle star *Ophiocoma wendtii* revealed this organism's use of magnesium as a toughening and strengthening strategy.[43] Synchrotron X-ray diffraction patterns taken from the lenses appeared to be composed of a Mg-calcite single phase, and chemical analysis showed that the total amount of incorporated Mg was about 15 mol% (i.e., [Mg]/([Ca]+[Mg]) = 15 mol%). On the other hand, analysis of the structure by means of aberration-corrected high-resolution transmission electron microscopy (HRTEM) revealed nano-inclusions of a somewhat lighter phase that were coherent with the matrix (**Figure 1**a,b). Further analysis, including mapping by Mg-EFTEM (energy-filtered transmission electron microscope), PEEM (photoemission electron microscopy) and ToF-SIMS (time-of-flight secondary ion mass spectrometry), showed that these inclusions were relatively richer in Mg. Moreover, upon heating, these nano-inclusions grew (from 5 to 20−40 nm) lost their coherence with the matrix (Figure 1c). Concurrently obtained diffractograms revealed the appearance of an apparently new Mg-rich calcite phase. Taken together, the results indicated that these inclusions are composed of magnesium-rich calcite, which is coherently included in a low Mg-calcite matrix.

Since Mg is smaller than Ca, the richer the calcite is in Mg, the smaller its lattice parameters. Thus, the lattice parameter is smaller in Mg-rich inclusions and larger if the magnesium-calcite matrix is poor. However, the fact that these inclusions are coherently present in the lenses implies that the entire material—prior to heating—displays the same lattice parameter and that the matrix is in a pre-compressed state. This is why the lattice parameter of the nanophase does not differ from that of the matrix prior to heating; moreover, if the diffraction peaks are plotted on a logarithmic scale, a broad hump that is characteristic of the nanophase can indeed be observed beneath each diffraction peak (Figure 1d).

Furthermore, synchrotron submicron scanning diffraction revealed the existence of layers with alternating d-spacings. These layers were also visible by nanotomography as alternating layers of varying density. These alternating d-spacings and densities probably arise from varying concentrations of Mg-rich nano-inclusions in the lenses. Thus, pre-compression of the material not only occurs in the entire bulk but also exists at different concentrations in alternating layers. This induces an even more pronounced enhancement of the mechanical properties since the layers lead to further crack deflection, as we observed in nanotomography of a cracked lens. Quantitative line profile analysis combined with TEM imaging showed that the nano-inclusions are 5 nm in size, correspond to 8 vol% of the material and are under an elastic tensile strain of 2.3%, whereas the matrix shows an elastic compressive strain of −0.1%, corresponding to a compressive stress of −170 MPa. This compressive stress in the lens results in an enhancement that is 1.36 times the strength of geological calcite. Moreover, our nano-indentation experiments indicated a fracture toughness 2.21 times that of geological calcite. Following loss of coherency (i.e. lattice relaxation), toughness and hardness of heated crystals were reduced. We believe that the formation of these nano-inclusions is initiated concurrently with the crystallization of an amorphous precursor supersaturated with Mg. Interestingly, the nano-inclusions in the Mg-rich calcite lenses resembled the well-known *Guinier-Preston* (GP) zones in metallurgy, [44] which are obtained by rapid quenching from high temperatures. The brittle star therefore exemplifies the ability of biominerals to form elaborate structures under ambient conditions, in contrast to the harsh conditions needed to obtain the man-made equivalent.

## 3. Residual strain caused by interaction with organic tissue

### 3.1. Compressive strains at the apatite/collagen interfaces

There are fascinating lessons to be learned from the constituent materials of teeth, owing to the fact that once these biominerals are formed they can resist repetitive mechanical loading from mastication for decades, without remodeling or self-healing. The main mass of the tooth consists of dentin, which is located below the enamel and surrounds the entire pulp. Dentin is composed of cell extensions called dental tubules, which emerge from the pulp through the dentin. In the peritububular dentin (PTD) the tubules are bordered by hydroxyapatite mineral tablets, and are surrounded by the matrix, known as the intertubular dentin (ITD). The ITD is made of collagen fibers, which are mineralized with hydroxyapatite and are orthogonal to the tubules[45] (**Figure 2**). Owing to the fibers orientation, cracks will propagate between them rather than across them towards the pulp.[45] The mineral tablets are approximately 3 nm thick.[46] They are randomly oriented in the PTD[47] but their c-axes are orientated mainly along the collagen fiber axis in the ITD.[48, 49] Measurement of the *c*-lattice parameters of the dentin hydroxyapatite crystals by X-ray diffraction nanotomography[50, 51] revealed that the *c*-lattice parameter of the crystals whose *c*-axis is aligned with the collagen fibers' axis are 0.08% smaller than those from the crystals having other orientations. Moreover, this difference disappeared after heating to 250°C. Therefore, the crystals aligned with the organic fibers are under compressive strains, that are associated with a compressive stress of 90 MPa. This difference in the *c*-lattice parameter and its disappearance after heating are attributed to the organic/inorganic interactions between the collagen and the apatite that cause the crystals to be compressed, but whose effect disappears when the organic matter is destroyed at high temperatures. Interestingly, the crystals under compression are the ones that are orthogonal to the tubules (the ones that are oriented along the collagen fibers in the ITD), so they are the ones responsible for directing the cracks orthogonally to the tubules rather than towards the pulp. The fact that they are under compressive strains makes them stronger and tougher and as a result more efficient in protecting the pulp.[45, 50, 51]

Nanoparticles of carbonated hydroxyapatite minerals attached and aligned with collagen fibrils are a common design in the bone family.[52] Evidence of compressive stresses in the mineral bound to the collagen was also reported in the case of bone[53-56] and was estimated to be between −160 and 250 MPa along the c-axis.[53]

Collagen fibrils in the bone materials are hydrated, and water removal was proved to result in collagen shrinkage,[57] which can induce high compressive strains in the attached mineral (Figure 2). In the case of dentine, water removal of the collagen fibrils (by drying) was shown to cause an increase of the compressive stress in the mineral of up to 300 MPa[51] (corresponding to 0.3% compressive strains). Similarly, water removal in turkey leg tendon caused localized strains of 1% in apatite crystals, corresponding to compressive stresses reaching 800 MPa.[58] The fact that collagen shrinkage can be translated to apatite compression to such a remarkable extent highlights the intimacy of the organic/inorganic interactions in this system.

Clues of possible residual compressive strains were also found at the surface of the fluoroapatite crystals of parrotfish teeth enameloid:[59] Strain maps obtained by X-ray microdiffraction revealed that the crystals showed small compressive strains of $4.10^{-4}$ along the *c*-axis near the teeth surface and that the absolute strain decreased and disappeared deeper in the material. The origin of this strain gradient is unknown.

### 3.2. Strained multi-layered structure of nacre

Bivalve shells consist of two bent shell valves pulled together by the adductor muscle. The shell is composed of nacre, in which calcium carbonate lamellas are connected via organic substances at their interfaces ('brick and mortar' design).[1] Due to the bending, the inner surface is compressed whereas the outer surface is in tension. These strains become imprinted in the mineral as the shell grows under the ligament stress. We analyzed the residual strains existing in the bent valve of *Perna canaliculus* shells, using a strain gauge attached to the inner surface, while its outer surface was being progressively etched (**Figure 3**a).[60, 61] We observed that etching of the outer surface resulted in a decrease of the bending curvature, corresponding to strain release. However, the results of the experiment showed that the strain release started only after 700 μm of the outer shell had already been etched, and it then increased with the amount of material etched (up to a measured strain release of 0.17%) (Figure 3b). These observations rebut previous hypothesis that the strains in the bent valves are attributable to the forces applied by the adductor muscle when closing the valves.[62] Indeed, if this were the case, the strain release would decrease with the amount of etched material.[60] Moreover, the increase in the released stress with the depth of etched material exhibits notable oscillations. Therefore, this strain release behavior suggests a structural reason for the stresses found in the valve. Our study revealed that this behavior stems from the multilayered structure of nacre, which is composed of alternating sublayers of ceramic and organic matter that display structural mismatching at their interface.[63] The authors developed a model showing that in a multilayered structure the increase in the strain release as well as the oscillations can be anticipated, since the multilayered structure by definition presents spatial periodic variations. In addition to the existence of strains at the interfaces, the high curvature of the valve is made possible by the fact that the strain varies through the valve depth. The thickness of the ceramic lamella indeed decreases with the valve depth, being thicker near the outer surface and thinner near the inner surface, (as demonstrated experimentally from HRSEM images), causing a depth-dependent strain gradient. This model corroborates the observed in-depth increase of microstrain fluctuations that was related to the in-depth increase of organic-inorganic interfaces.[64] Therefore, the model agrees with the experimental data and shows that both the high curvature of the valve and the gradual strain release with the etching of the valve's outer surface result from the strains existing at the organic-inorganic interfaces. The strains imposed by the interfaces are due to structural mismatching between the organic and inorganic layers and are depth dependent.

## 5. Conclusion and outlook

Residual strains occur in numerous biogenic crystals, including calcite and aragonite mollusk shells, calcitic lenses, hydroxyapatite. and fluoroapatite. We showed here that these strains exist as a result of different mechanisms. One such mechanism is the integration of organic or inorganic inclusions in a calcium carbonate lattice crystallized from an amorphous phase. Another mechanism is the utilization of strong organic or inorganic interaction at organic/inorganic interfaces, causing strains in the mineral as a result of the structural mismatch between the phases.

The list of biogenic systems with reported residual strains is still relatively short, possibly because these systems can be difficult to detect. Lattice distortions arising from chemical variability or from the presence of impurities often need to be recognized and taken into account, and lattice strains arising from different mechanisms can counteract one other. Moreover, most of the attention up to now has been directed to structural features at larger scales. However, we feel that it is important to take into account the possible existence of residual strains when investigating the multi-scale structure of biominerals. We believe that other cases and mechanisms of internal strains are yet to be discovered, and that their elucidation will be of interest for the full understanding of biomineral structures and the design of novel bio-inspired engineering solutions.

**Acknowledgements**
The present research received funding from the European Research Council under the European Union's Seventh Framework Program (FP/2007–2013)/ERC Grant Agreement No. 336077 and from the European Union Horizon 2020 research and innovation program under the Marie Skłodowska-Curie grant agreement no. 642976-NanoHeal Project.

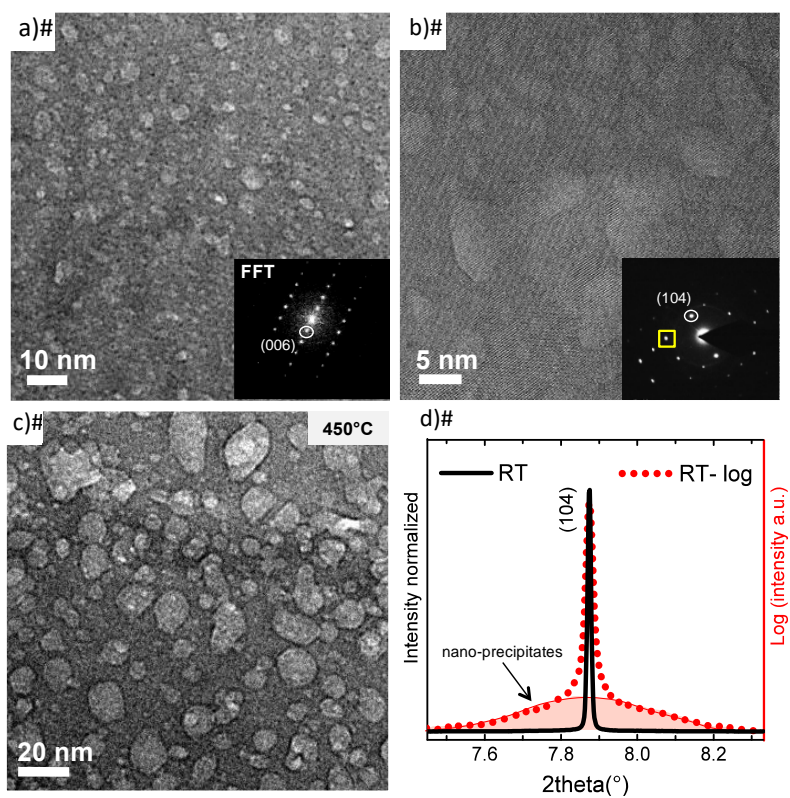

**Figure 1**. a) Bright-field HRTEM image of a thin section from a lens, revealing brighter nanodomains, although the fast Fourier transform (FFT) pattern is that of a single crystal. b) Higher magnification of an area in a) showing continuous lattice fringes across the nanodomains, indicating that they have coherent interfaces with the lattice. The inset shows an electron diffraction image from this area. c) Bright-field HRTEM image obtained at 450°C, revealing the temperature-dependent growth of the nanodomains. d) (104) diffraction peak of a powdered dorsal arm plate sample at room temperature collected at a wavelength of 0.4106 Å, comparing linear (black) and logarithmic (red) intensity scales and revealing the presence of nanodomains at the base of the diffraction peak. Reproduced with permission.[43] Copyrights 2018, AAAS.

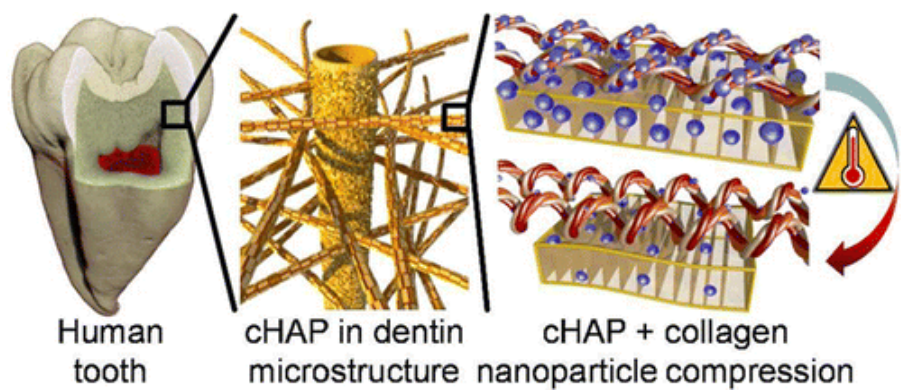

**Figure 2**. Schematic representations of a human tooth, of the microstructure of dentin represented by a tubule and orthogonal mineralized collagen fibrils; and of the effect of water removal on the apatite crystals attached to the collagen fibril: the collagen molecules contract,

causing a compression of the apatite crystals. The vertical planes represent the (002) planes of apatite. Reproduced with permission.[51] Copyright 2018, American Chemical Society.

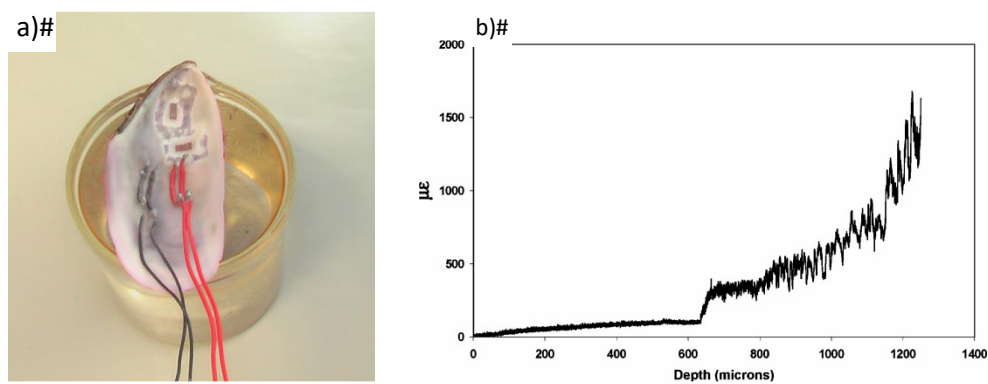

**Figure 3**. a) Picture of the experimental set-up: strain gauges attached to the inner surface of a *Perna canaliculus* shell measure the strain released by the etching of its outer surface. b) Strain release (in the unit of $10^{-6}$) versus depth of etching, showing an increase of the strain release and pronounced oscillations beneath its inner surface. Reproduced with permission.[60] Copyright 2009, Wiley-VCH.